\documentclass[aps,prx,twocolumn,superscriptaddress, showpacs]{revtex4-2}

\usepackage{graphicx}
\usepackage{dcolumn}
\usepackage{bm}
\usepackage{color}
\usepackage{ulem}
\usepackage[usenames,dvipsnames,svgnames,table]{xcolor}

\usepackage{amsfonts}       % ajoute des polices mathématique
\usepackage{amsmath}        % ajoute des environnements mathématiques
\usepackage{bm}             % ajoute des caractères grecs en gras
\usepackage{mathrsfs}       % ajoute une meilleure police calligraphique pour certains symboles
\usepackage{setspace}       % gère l'interligne
\usepackage{microtype}     % permet de ne pas déborder à droite en bout de ligne
\usepackage{enumerate}% http://ctan.org/pkg/enumerate
\usepackage{soul}      % for highlighting
\usepackage{xspace}
\usepackage[unicode=true,
  linktocpage,
  linkbordercolor={0.5 0.5 1},
  citebordercolor={0.5 1 0.5},
  linkcolor=blue]{hyperref}
\usepackage{changes}
\usepackage{makecell}

\newcommand{\Tstar}{$T^{\star}$\xspace}
\newcommand{\ra}{$\rho_{\rm a}$\xspace}
\newcommand{\rb}{$\rho_{\rm b}$\xspace}
\newcommand{\Sa}{$S_{\rm a}$\xspace}
\newcommand{\Sb}{$S_{\rm b}$\xspace}

\begin{document}

%\preprint{}

%-----------------------------------------%
%############### Title ###################%
%-----------------------------------------%

\title{No nematicity at the onset temperature of the pseudogap phase\\
in the cuprate superconductor YBa$_{\rm 2}$Cu$_{\rm 3}$O$_{\rm y}$}

%\title{No additional charge nematicity at the pseudogap phase temperature onset in YBa$_{\rm 2}$Cu$_{\rm 3}$O$_{\rm y}$}
% \title{Absence of charge nematicity\\ at the pseudogap phase temperature onset in YBa$_{\rm 2}$Cu$_{\rm 3}$O$_{\rm y}$}

%-----------------------------------------%
%############### Authors #################%
%-----------------------------------------%



\author{G.~Grissonnanche}
\email{gael.grissonnanche@cornell.edu}
\affiliation{Institut quantique, D\'{e}partement de physique  \&  RQMP, Universit\'{e} de Sherbrooke, Sherbrooke,  Qu\'{e}bec, Canada}
\affiliation{Laboratory of Atomic and Solid State Physics, Cornell University, Ithaca, NY, USA}
\affiliation{Kavli Institute at Cornell for Nanoscale Science, Ithaca, NY, USA}

\author{O.~Cyr-Choini\`{e}re}
\affiliation{Institut quantique, D\'{e}partement de physique  \&  RQMP, Universit\'{e} de Sherbrooke, Sherbrooke,  Qu\'{e}bec, Canada}

\author{J.~Day}
\affiliation{Department of Physics and Astronomy, University of British Columbia, Vancouver, British Columbia, Canada}

\author{R.~Liang}
\affiliation{Department of Physics and Astronomy, University of British Columbia, Vancouver, British Columbia, Canada}

\author{D.~A.~Bonn}
\affiliation{Department of Physics and Astronomy, University of British Columbia, Vancouver, British Columbia, Canada}

\author{W.~N.~Hardy}
\affiliation{Department of Physics and Astronomy, University of British Columbia, Vancouver, British Columbia, Canada}

\author{N.~Doiron-Leyraud}
\affiliation{Institut quantique, D\'{e}partement de physique  \&  RQMP, Universit\'{e} de Sherbrooke, Sherbrooke,  Qu\'{e}bec, Canada}

\author{L.~Taillefer}
\email{louis.taillefer@usherbrooke.ca}
%\altaffiliation{Correspondence should be addressed to L.T. (e-mail: louis.taillefer@usherbrooke.ca)}
\affiliation{Institut quantique, D\'{e}partement de physique  \&  RQMP, Universit\'{e} de Sherbrooke, Sherbrooke,  Qu\'{e}bec, Canada}
\affiliation{Canadian Institute for Advanced Research, Toronto, Ontario, Canada}


\date{\today}

% %-----------------------------------------%
% %############### Abstract ################%
% %-----------------------------------------%

\begin{abstract}

Electronic nematicity is the spontaneous loss of rotational symmetry in a metal, without breaking translational symmetry.
%
%In some theoretical models and for some materials, a link is proposed between nematicity and unconventional superconductivity.
%
In the cuprate superconductors, there is experimental evidence for nematicity, but its origin remains unclear.
Here we investigate the onset of nematicity in the transport of charge by means of electric and thermoelectric measurements in underdoped YBa$_{\rm 2}$Cu$_{\rm 3}$O$_{\rm y}$, performed by passing the current (electrical or thermal) first along the $a$ axis then the $b$ axis of the orthorhombic structure in the same crystal, with a hole doping $p = 0.12$.
Upon cooling, we observe no additional in-plane anisotropy -- beyond the background anisotropy due to the CuO chains --
in either the resistivity $\rho$ or the Seebeck coefficient $S$ as the temperature $T^{\star}$~for the onset of the pseudogap phase is crossed.
We conclude that the pseudogap phase of cuprates is not nematic.
However, at temperatures much lower than $T^{\star}$, a strong additional anisotropy is observed,
most clearly in the Peltier coefficient $\alpha = S / \rho$.
%in both transport coefficients,
We interpret it as nematicity associated with the development of charge order.

\end{abstract}

%-----------------------------------------%
%############### PACS ####################%
%-----------------------------------------%

\pacs{74.72.Gh, 74.25.Dw, 74.25.F-}

% 74.72.Gh  Hole-doped cuprate superconductors
% 74.25.Dw  Phase diagrams superconductivity
% 74.25.F-  Transport properties

%-----------------------------------------%
%############### Keywords ################%
%-----------------------------------------%

%\keywords{}

%-----------------------------------------%
%############### Make Title ##############%
%-----------------------------------------%

\maketitle

%-----------------------------------------%
%############### MAIN ####################%
%-----------------------------------------%

% References should be done using the \cite,\onlinecite, \ref, and \label commands

%>>>>>>>>>>>>>>>>>>>>>>>>>>>>>>>>>>>>>>>>>>>>>>>>>>>>>>>>>>>>>>>>>>>>>>>>>>>>>>>>>>>>>>>>>>>>>>>>>>
\section{INTRODUCTION}

The way in which strong electronic correlations give rise to high temperature superconductivity remains a mystery.
 Hole-doped cuprates possess the highest superconducting transition temperatures of all unconventional superconductors. In these quantum materials, superconductivity coexists with the pseudogap phase -- one of the most studied enigma of strongly correlated materials for the role it might play in the mechanism of electron pairing~\cite{Keimer2015From,proust_remarkable_2019}.

\begin{figure}[h!]
\centering
\includegraphics[width=0.45\textwidth]{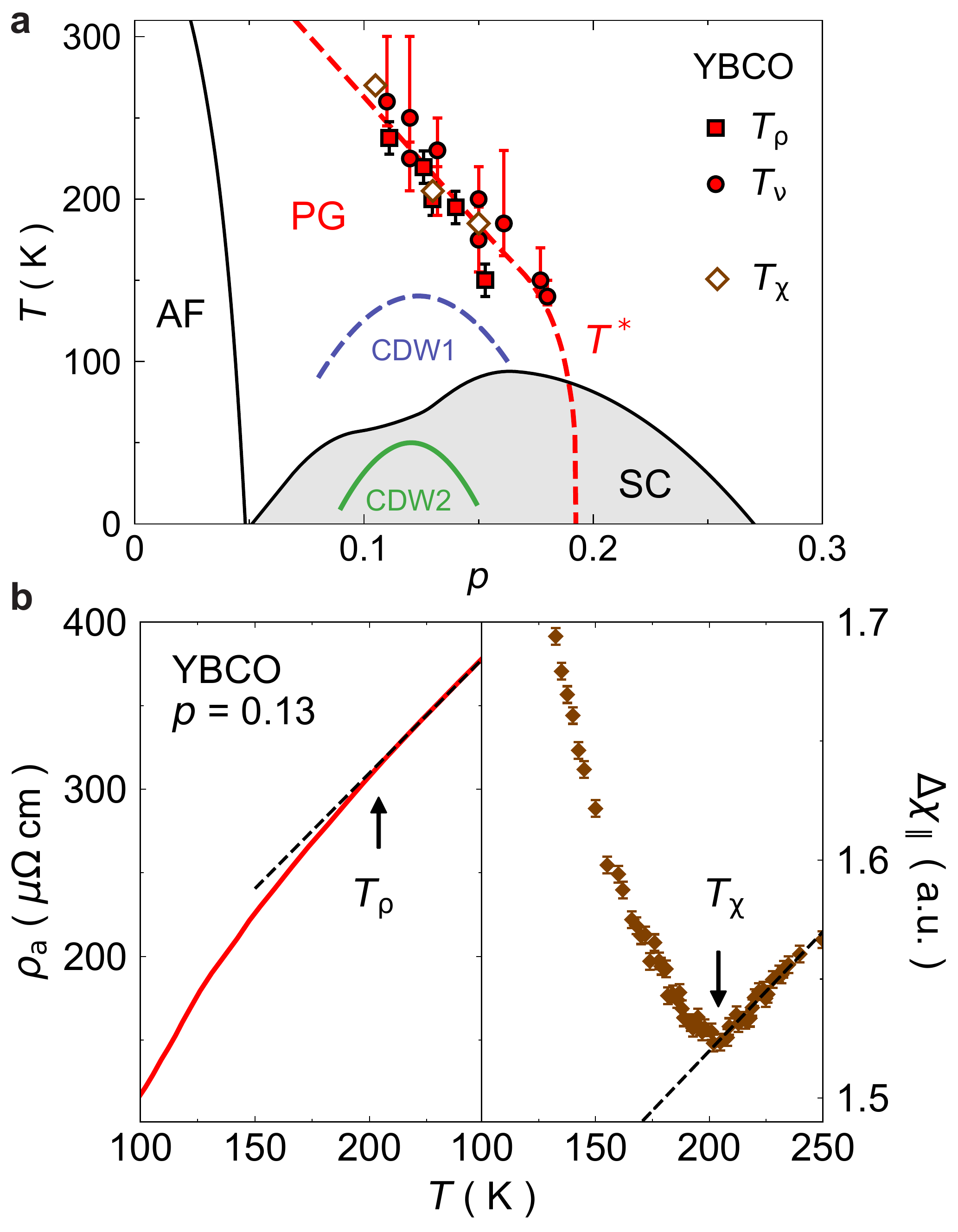}
\caption{
(\textbf{a})
Temperature vs doping phase diagram of YBCO, showing
the antiferromagnetic phase (AF),
the dome of superconductivity (SC)
and the pseudogap phase (PG),
as well as the regions of short-range charge-density-wave correlations (CDW1) and long-range charge order (CDW2).
The red dashed line marks \Tstar, the onset temperature of the pseudogap phase,
defined by the drop in either the resistivity ($T_{\rm \rho}$, full red squares) or
the Nernst coefficient ($T_{\rm \nu}$, full red circles)~\cite{cyr-choiniere_pseudogap_2018}.
$T_{\rm \chi}$ marks the onset of anisotropy in torque measurements (open brown diamonds~\cite{Sato2017Thermodynamic}).
%$T_{\rm \nu}$ shows the signature of the \Tstar in the Nernst;
%
(\textbf{b})
{\it Left panel}:
temperature dependence of the $a$-axis resistivity of YBCO at $p=0.13$ \cite{ando_electronic_2004}.
The deviation from $T$-linear resistivity (dashed line) below $T_{\rm \rho}$ defines \Tstar.
{\it Right panel}:
in-plane anisotropy of the magnetic susceptibility $\chi$, plotted as $\Delta\chi_{\parallel}$ vs $T$,
as measured by torque on YBCO at $p=0.13$~\cite{Sato2017Thermodynamic}.
$T_{\rm \chi}$ marks a sharp signature in the data, found to coincide with \Tstar.
}
\label{fig:Tstar_definition}
\end{figure}

The pseudogap phase appears below a temperature \Tstar~(defined by a change in either the electrical resistivity or the Nernst coefficient~\cite{cyr-choiniere_pseudogap_2018}) that ends inside the superconducting dome (Fig.~\ref{fig:Tstar_definition}a).
Among the many reports of broken symmetries below \Tstar~\cite{fauque_magnetic_2006,xia_polar_2008,Zhao2016Global,lubashevsky_optical_2014}, the loss of rotational symmetry without breaking translational symmetry -- known as nematicity  -- has been a recurring theme not only in cuprates, but also in other unconventional superconductors~\cite{lawler_intra-unit-cell_2010,chu_2012,cao_2021}, 
with an unequivocal demonstration in iron pnictide superconductors based on the anisotropy of the resistivity \cite{chu_2012}.

In cuprates, there is strong evidence of nematicity deep inside the pseudogap phase~\cite{ando_electrical_2002,hinkov2007,cyr-choiniere_two_2015} and some studies -- based on Nernst \cite{Daou2010Broken}, magnetic torque \cite{Sato2017Thermodynamic} and elastoresistance \cite{ishida_2020} measurements -- report the onset of nematicity at \Tstar~(Fig.~\ref{fig:Tstar_definition}).
These studies indicate that the pseudogap phase is inherently nematic, which in turn suggests that nematicity may play a role in boosting superconductivity \cite{maier_2014, lederer_2015}.

%#################### Figure 1 ####################%
%##################################################%

In the cuprate YBa$_{\rm 2}$Cu$_{\rm 3}$O$_{\rm y}$ (YBCO), magnetic torque experiments \cite{Sato2017Thermodynamic} report the sudden onset of broken rotational symmetry at a temperature $T_{\rm \chi}$ that coincides with $T^{\star}$ (Fig.~\ref{fig:Tstar_definition}b), 
but it is unclear whether this nematicity is a property of the electron fluid.
Nernst experiments \cite{Daou2010Broken} show the gradual onset of an additional anisotropy -- between the $a$ and $b$ crystallographic directions of the orthorhombic crystal structure -- at a temperature $T_{\rm \nu}$ that also coincides with $T^{\star}$.
% \gael{(Fig.~\ref{fig:nernst})}.
%
This points towards the onset of nematicity, at \Tstar, in the electron fluid that flows in the CuO$_2$ planes.
However, the Nernst coefficient $\nu$ is a composite of four electric and thermoelectric, longitudinal and transverse, transport coefficients:
\begin{equation}
\nu_{\rm i} = R_{\rm H}\alpha_{\rm i} - \frac{\alpha_{\rm ij}\rho_{\rm i}}{B}~~~~,
\label{eq:nernst}
\end{equation}
where $\nu_{\rm i}$ is the Nernst coefficient for a current along the $i$ direction ($i, j=a$ or $b$), $R_{\rm H}$ is the Hall coefficient, $\alpha_{\rm i}$ and $\alpha_{\rm ij}$ are the longitudinal and transverse Peltier coefficients, and $\rho_{\rm i}$ is the resistivity.
This makes the Nernst coefficient difficult to interpret, since a change in any of the four constituent coefficients can create a collateral anisotropy between $\nu_{\rm a}$ and $\nu_{\rm b}$ that would not be an expression of nematicity.
This calls for the measurement of a pure longitudinal transport coefficient in order to settle the existence of nematicity
occurring at \Tstar.

In this Article, we investigate the nematicity in the charge transport properties of YBCO, at a hole doping $p = 0.12$.
At this doping, the pseudogap temperature is $T^{\star} = 220\pm10$~K~\cite{ando_electronic_2004}, the short-range charge density wave correlations (CDW1) onset at $T_{\rm CDW1} = 140\pm10$~K~\cite{chang_direct_2012}, and the charge density wave order becomes long-range (in a field larger than $H=15$~T) and unidirectional below $T_{\rm CDW2} = 47\pm1$~K~\cite{chang_magnetic_2016} (Fig.~\ref{fig:Tstar_definition}a).

To reliably measure the transport anisotropy between the $a$ and $b$ directions of the orthorhombic structure in YBCO, the longitudinal electric and thermoelectric transport coefficients -- the resistivity $\rho$ and the Seebeck coefficient $S$ -- were measured on the same sample with the same contacts, before and after inverting the $a$ and $b$ directions in the single crystal.
%
%however the crystallographic direction of the sample was switch during the experiment.
%
Following this method, we detected no additional anisotropy in either transport coefficient upon cooling through \Tstar. 
%This shows that the nematicity reported at \Tstar~does not reside in the charge sector.
%
%Next, we revisit the onset of anisotropy in the Nernst at \Tstar and propose that it can be explained by the drop in the resistivity at the same temperature. We conclude the absence of charge nematicity at the onset of the pseudogap phase of cuprates.
%
As explained below, the onset of nematicity at \Tstar~reported previously from Nernst measurements is an artefact of the composite nature of that transport coefficient.
We conclude that the pseudogap phase is not inherently nematic in its metallic properties.
However, at this doping, there is indeed a strong nematicity that emerges deep inside the pseudogap phase, 
in tandem with the development of charge order at low temperature.

%>>>>>>>>>>>>>>>>>>>>>>>>>>>>>>>>>>>>>>>>>>>>>>>>>>>>>>>>>>>>>>>>>>>>>>>>>>>>>>>>>>>>>>>>>>>>>>>>>>

\section{METHODS}

\textbf{Material}.
To study nematicity in a tetragonal crystal,
four-fold rotational symmetry must ideally be broken on a macroscopic scale in order for any nematic domains, if present, to all be aligned. This is the necessary condition for a nematic monodomain to inform and be amenable to transport anisotropy studies.
One way to achieve this is to apply uniaxial pressure to force domains to align in one direction,
as done in iron pnictides~\cite{chu_2012}.
The cuprate material YBCO has been a favorite playground for the study of nematicity
because its orthorhombic structure will naturally align nematic domains.
Indeed, by detwinning a single crystal of YBCO, so that it becomes a single structural monodomain,
 the entire sample hosts a preferential direction.
In other words, the orthorhombic structure then becomes an advantage, for it is what actually allows us to measure nematicity. Indeed, the orthorhombicity acts like the uniaxial pressure applied in the pioneer elastoresistivity experiments performed on iron-based superconductors~\cite{chu_2012}, except that for YBCO, this pressure is internal and not external.
However,
the orthorhombic structure of YBCO comes with unidirectional CuO chains that run along the $b$ axis of the crystal,
and these chains are conductive.
They therefore impose a background anisotropy in the transport properties that one needs to carefully take into account.
% induce a uniaxial ``chemical pressure'' that can help align any type of domain.\tabularnewline
%
%Nevertheless, the CuO chains also induce an orthorhombic structural distortion that starts above room temperature.
%
%In addition, the CuO chains contribute to the transport of charge and heat along the $b$-axis, complicating the study of the nematicity in the CuO$_2$ planes where exotic physics develops.
%
%Although the anisotropy associated to chains does not correlate with the in-plane nematicity observed in YBCO \cite{ando_electrical_2002,Daou2010Broken,cyr-choiniere_two_2015}, this approach calls for prudence as the chains may mask the onset of additional anisotropy that could appear at \Tstar and would be synonym of nematicity.

\textbf{Samples}.
A single crystal of YBa$_{\rm 2}$Cu$_{\rm 3}$O$_{\rm y}$ (YBCO) was prepared at the University of British Columbia by flux growth \cite{liang_growth_2012}. The detwinned sample is
an uncut, unpolished thin platelet, with gold evaporated contacts (of resistance $< 1 \Omega$). Its hole concentration (doping) $p$ is determined from the superconducting transition temperature $T_{\rm c}$ \cite{liang_evaluation_2006}, defined as the temperature below which the zero-field resistance is zero. A high degree of oxygen order was achieved with $p=0.12$ ($y = 6.67$, ortho-VIII order) and $T_{\rm c} = 65.5$~K.
To be able to compare with precision the measurements between the $a$- and $b$-axis directions, we had to get rid of the uncertainties associated with the geometric factors inherent to transport experiments --- typically $\approx 15\%$ uncertainty because contacts have a certain width.
To achieve this, the experiment was performed in a two-stage process.
First, gold contacts were deposited on the surface of the sample in order to measure the $a$-axis transport coefficients (see Fig. \ref{fig:resistivity}a).
Then, the same sample with the same contacts was detwinned in the other crystallographic direction, which means $a$ and $b$ were rotated by 90$^\circ$,
allowing us to measure the $b$-axis transport coefficients with the same set of contacts.

\textbf{Measurements}.
We measured the resistivity and Seebeck coefficient on the same YBCO $p=0.12$ sample for $a$- and $b$-axis orientations of the crystal. Silver wires glued with silver paste on the gold contacts of the sample were used for both measurements. 
%The experiment was performed in a Quantum Design PPMS.
%
The electrical resistivity $\rho$ was measured by sending an electric current along $x$, and measuring the associated voltage difference along $x$,
where $x$ was first along the $a$ axis and then along the $b$ axis, using the exact same contacts.
The Seebeck coefficient $S$ was measured by sending a heat current along $x$, and measuring the associated voltage difference and temperature difference along $x$, so that $S = \Delta V / \Delta T$, where again $x$ was first along the $a$ axis and then along the $b$ axis.
Note that a measurement of $S$ involves two types of contacts -- electrical and thermal --
whose effective position along the length of the sample will in general be slightly different (although measured using the same gold pads),
resulting in an uncertainty on $S$ given roughly by contact width over contact separation, so about 10\%.
This will introduce a small uncertainty on the Peltier coefficient $\alpha$, which invloves taking the ratio
of $S$ over $\rho$.
%
%
%\textcolor{red}{
%(Note that the very small anisotropy in $\alpha$ above 140~K (inset of Fig.~\ref{fig:Seebeck}c))
%which amounts to $\Delta \alpha / \alpha \simeq 5 \%$,
%is within the uncertainty of this composite measurement, 
%which involves taking the ratio of two separate quantities, $S$ and $\rho$,
%given that $S$ involves two types of contact -- electrical and thermal --  effective position along the length of the sample will in general be slightly different -- although measured from the same gold pads,
%resulting in an uncertainty on $S$ given roughly by contact width over contact separation, so about 10\%).
%}
%

A magnetic field of 16~T was applied along the $c$ axis, perpendicular to the CuO$_2$ planes, in order to partially suppress superconductivity
and therefore extend the measurements to lower temperature.

%>>>>>>>>>>>>>>>>>>>>>>>>>>>>>>>>>>>>>>>>>>>>>>>>>>>>>>>>>>>>>>>>>>>>>>>>>>>>>>>>>>>>>>>>>>>>>>>>>>
\section{RESULTS}

To detect nematicity in YBCO,
we measured the anisotropy of the resistivity and the Seebeck coefficient -- two longitudinal probes of charge transport --
between the $a$ and $b$ directions in our orthorhombic (monodomain) crystal.

%#################### Figure 2 ####################%
\begin{figure}[t!]
\centering
\includegraphics[width=0.395\textwidth]{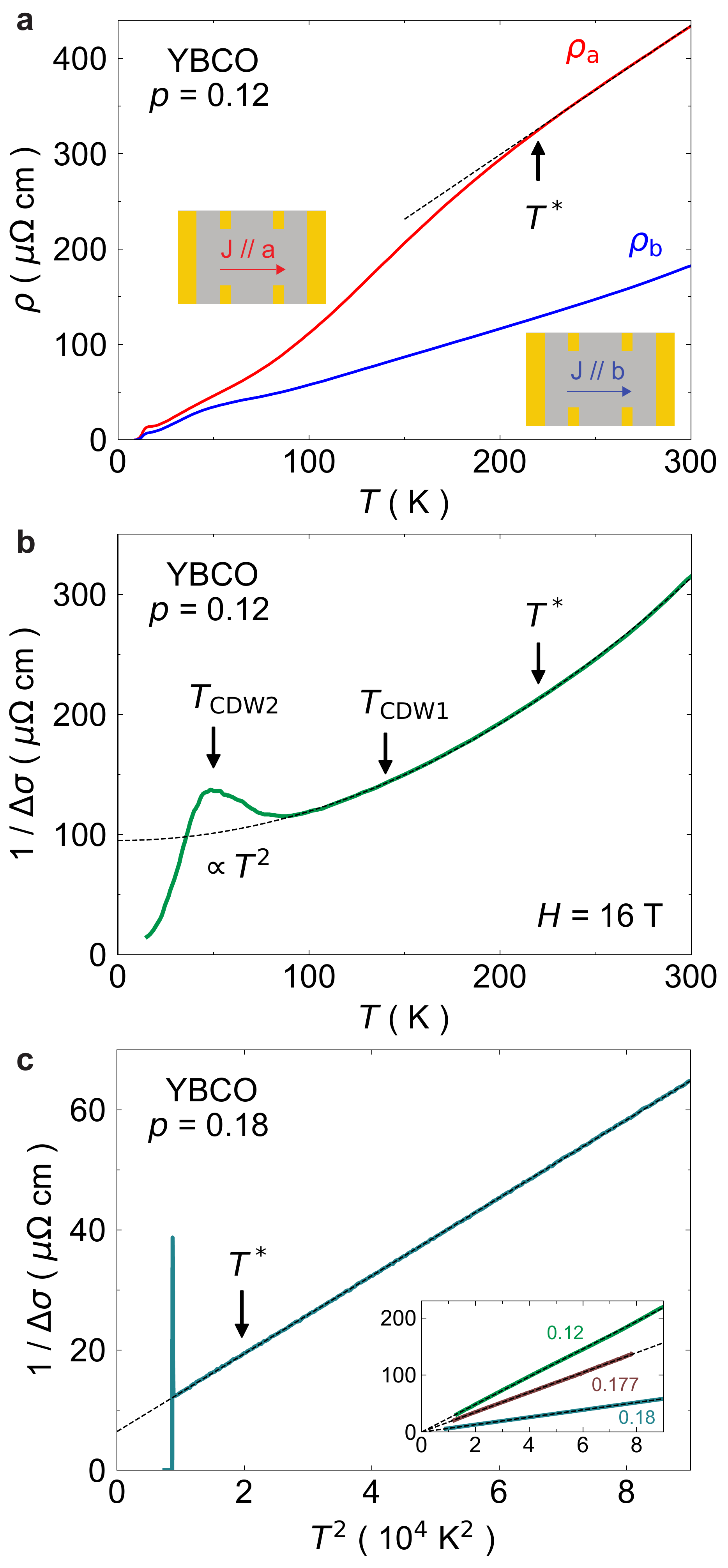}
\caption{
In-plane resistivity of YBCO at $p=0.12$, in a magnetic field of $H=16$~T.
(\textbf{a})
Temperature dependence of the resistivity along the $a$ axis ($\rho_a$, red) and $b$ axis ($\rho_b$, blue) of the orthorhombic crystal structure.
\Tstar~is defined as the departure in $\rho_a$ from the $T$-linear dependence at high temperature (black dashed line).
(This is called $T_{\rho}$ in Fig.~1, but \Tstar~hereafter.)
A sketch of the sample and contacts is shown for each current direction.
(\textbf{b})
Resistivity anisotropy, plotted as $1/\Delta\sigma = 1 / (1/\rho_{\rm b}-1/\rho_{\rm a})$ vs $T$.
The $T^2$ dependence of $1/\Delta\sigma$ (dashed line) coming from the CuO chains is seen to persist down to $T \approx 100$~K, showing that no nematicity develops below \Tstar=220~K.
The deviation below $\simeq 100$~K occurs inside the region of charge-density-wave ordering.
%the anisotropy in the charge sector is not impacted by the onset of the pseudogap phase neither by the short-range charge density wave (CDW1).
%
(\textbf{c})
$1/\Delta\sigma$ vs $T^2$ of YBCO at $p=0.18$ ($y=6.998$) (from ref.~\onlinecite{Daou2010Broken}).
The CuO chains are fully oxygenated and maximally conductive at $p=0.18$ and follow a $T^2$ dependence down to $T_{\rm c}$.
 {\it Inset}:
 chain resistivity (with $T=0$ value subtracted),
 at $p=0.12$ (green, this work),
 $p = 0.177$ (violet, \cite{Daou2010Broken}),
 and
  $p = 0.18$ (blue, \cite{Daou2010Broken}),
Dashed lines are $T^2$ fits.}
\label{fig:resistivity}
\end{figure}
%##################################################%

%
%To search for an electronic anisotropy associated to nematicity at the onset of the pseudogap phase, we studied the anisotropy between the resistivities \ra and \rb.
%
\textbf{Resistivity}.
In YBCO, the $a$-axis resistivity \ra reflects purely the conductivity of the CuO$_2$ planes,
{\it i.e.} $1/\rho_{\rm a} = \sigma_{\rm plane}^{\rm a}$.
When the current is along $b$, however, there are two parallel channels of charge conduction:
the two-dimensional CuO$_2$ planes ($\sigma_{\rm plane}^{\rm b}$) and the one-dimensional CuO chains ($\sigma_{\rm chain}$),
so that $1/\rho_{\rm b} = \sigma_{\rm plane}^{\rm b} + \sigma_{\rm chain}$.
The large resulting anisotropy between \ra~and \rb~can clearly be seen in Fig.~\ref{fig:resistivity}a, 
in particular above \Tstar.

Fortunately, we can disentangle any anisotropy coming from the CuO$_2$ planes -- as expected in presence of nematicity -- from the background anisotropy due to the CuO chains by using the fact that the chain resistivity follows a perfectly defined $T^2$ dependence from 300~K down to at least 90~K, as established by measurements in YBCO at high doping~\cite{gagnon_t_1994,Daou2010Broken} and reproduced in Fig.~\ref{fig:resistivity}c.
In Fig. \ref{fig:resistivity}b, we examine the anisotropy of the conductivity, defined as the difference $\Delta\sigma = 1/\rho_{\rm b} - 1/\rho_{\rm a}$, by plotting our data as $1/\Delta\sigma$ vs $T$.
%
%while the physics of the planes gets polluted in the $b$-axis resistivity \rb by an extra conductivity coming from the one-dimensional CuO chains running along the $b$-axis and $1/\rho_{\rm b} = \sigma_{\rm plane}^{\rm b} + \sigma_{\rm chains}$. Therefore, already above the onset temperature of the pseudogap phase at \Tstar, \rb is lowered and a significant anisotropy between \ra and \rb can be observed because of the CuO chains (Fig. \ref{fig:resistivity}).

%To disentangle the parasitic CuO chains contribution to the conductivity any nematic behavior that would emerge from the CuO$_2$ planes, we examined the anisotropy of the conductivity between the $a$- and $b$-axis: $\Delta\sigma = 1/\rho_{\rm b} - 1/\rho_{\rm a}$
%
We observe that $1/\Delta\sigma$ follows a $T^2$ dependence down to $T\approx 100$~K,
entirely coming from the CuO chains.
The slope of the $T^2$ resistivity from the chains tends to decrease with increasing doping as showed in the inset of Fig.\ref{fig:resistivity}c, as naively expected for CuO chains more and more conductive as $p \rightarrow 0.18$ when fully oxygenated.
Although the resistivity of the CuO chains obeys a pristine $T^2$ dependence that remains to be theoretically understood, this allows us to subtract perfectly the chain contribution from the $b$ axis resistivity and access the pure anisotropy coming from the CuO$_2$ planes.
The fact that no additional anisotropy in $1/\Delta\sigma$ is detected at $p=0.12$ upon cooling below \Tstar~$= 220$~K shows that the pseudogap phase causes no nematicity to emerge in the electron fluid.
%
%We identify this dependence with the CuO chains that are known to behave like a one-dimensional metallic subsystem with a perfectly defined $T^2$-resistivity \cite{gagnon_t_1994,Daou2010Broken}. Not only $1/\Delta\sigma$ can be associated to the CuO chains only, but the absence of change in this $T^2$ behavior above and below \Tstar reveals that there is no additional anisotropy, therefore no onset of charge nematicity at the pseudogap phase temperature.

Note, however, that there is a clear deviation from the $T^2$ background that occurs below 100~K or so (Fig.~\ref{fig:resistivity}a).
We return to this feature in the discussion section.

\begin{figure}[t!]
\centering
\includegraphics[width=0.46\textwidth]{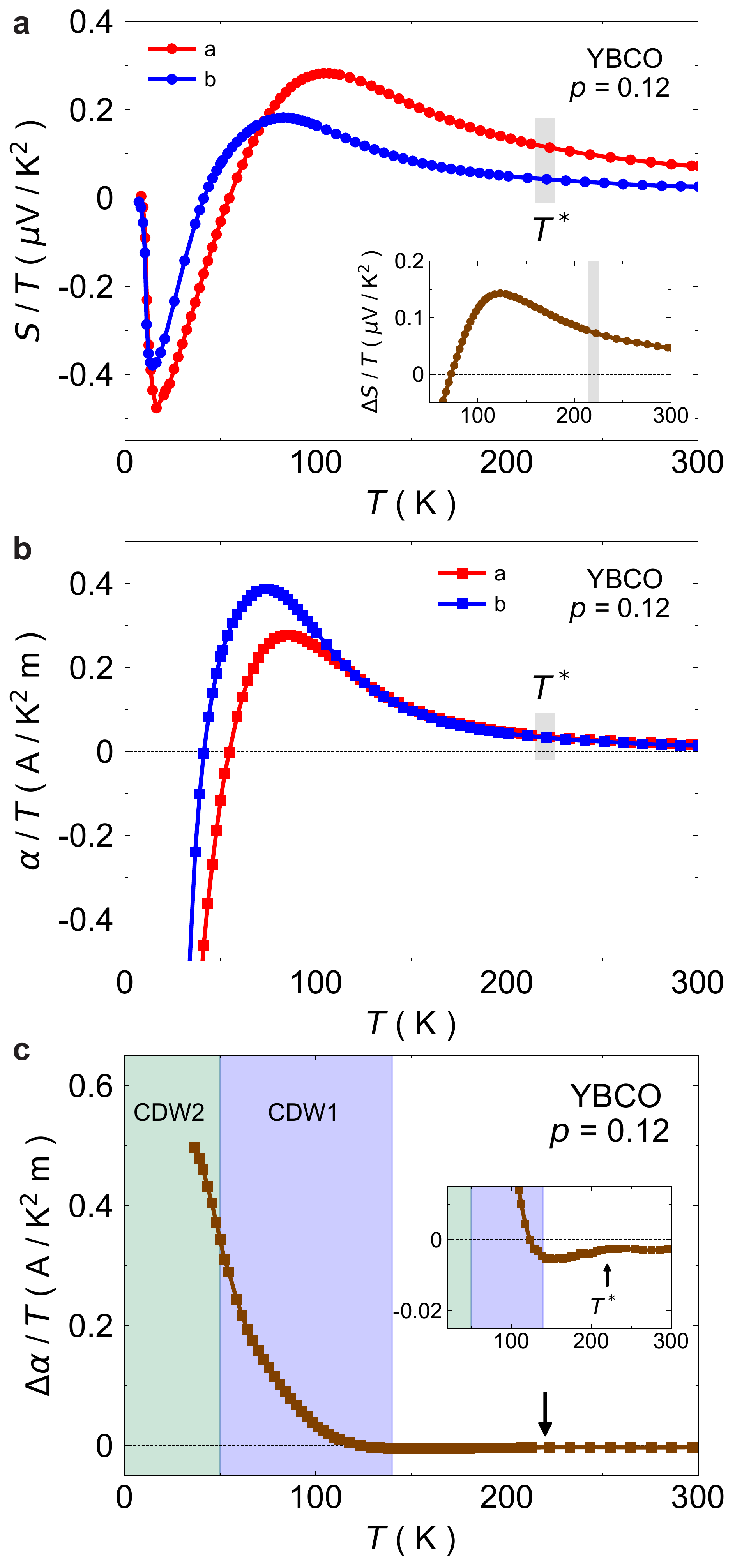}
\caption{
Thermoelectric transport coefficient of YBCO at $p=0.12$ in a magnetic field $H=16$~T.
(\textbf{a})
Temperature dependence of the Seebeck coefficient $S$ along the $a$ axis (red) and $b$ axis (blue).
The inset represents the difference $\Delta S / T = (S_{\rm a} - S_{\rm b}) / T$ vs $T$.
%The arrow indicates \Tstar.
$T^{\star}$ is marked by gray stripes.
(\textbf{b})
Peltier coefficient, $\alpha = S / \rho$, along $a$ axis (red) and $b$ axis (blue).
$T^{\star}$ is marked by a gray stripe.
(\textbf{c})
Anisotropy of the Peltier coefficient, plotted as $\Delta \alpha / T = (\alpha_{\rm b} - \alpha_{\rm a}) / T$ vs $T$.
The arrow marks \Tstar.
The blue shaded region indicates the regime of short-range two-dimensional charge-density-wave correlations (CDW1)
and the green shaded region the regime of long-range three-dimensional charge-density-wave order (CDW2).
{\it Inset}:
zoom at high temperature.}
\label{fig:Seebeck}
\end{figure}

\textbf{Seebeck coefficient}.
To confirm our findings from resistivity,
%the absence of charge nematicity onsetting at \Tstar, we
we measured the anisotropy of the Seebeck coefficient, another longitudinal coefficient.
Our data are shown in Fig.~\ref{fig:Seebeck}a, plotted as $S/T$ vs $T$.
A clear anisotropy is observed between \Sa~and \Sb, even above \Tstar.
%
%The same way the CuO chains contaminate the $b$-axis resistivity, a large anisotropy is present at any temperature between \Sa and \Sb (Fig.~\ref{fig:Seebeck}a).
In the inset, we plot the difference $\Delta S = S_{\rm a} - S_{\rm b}$,
which does not show any anomaly at \Tstar.
However, Seebeck is not an additive coefficient like the conductivity and we do not know what the background anisotropy from CuO chains should be in $S$. 
It makes
this statement on the absence of nematicity at \Tstar~less conclusive for the Seebeck coefficient than for the resistivity where the temperature dependence of the resistivity of CuO chains is known.
%indicating an absence of nematicity at the onset of the pseudogap phase.

%#################### Figure 3 ####################%
%##################################################%

\textbf{Peltier coefficient}.
To shed light on the anisotropy present in the Seebeck coefficient,
we turn to the Peltier coefficient $\alpha = S / \rho$ -- the fundamental thermoelectric transport coefficient.
The advantage
%to look at Peltier coefficient rather than the Seebeck coefficient
here is that $\alpha$ is additive while by definition $S$ is not.
In Fig.~\ref{fig:Seebeck}b, we plot $\alpha/T$ vs $T$ for both current directions.
We see, strikingly, that $\alpha$ is isotropic down to $ T \simeq 140$~K.
The difference $\Delta \alpha = \alpha_{\rm a} - \alpha_{\rm b} = S_{\rm a} / \rho_{\rm a} - S_{\rm b}/\rho_{\rm b}$, plotted as $\Delta \alpha / T$ vs $T$ in Fig.~\ref{fig:Seebeck}c, is essentially zero above that temperature (at least an order of magnitude smaller than the plane contribution), up to at least 300~K.
(Note that the very small anisotropy in $\alpha$ above 140~K (inset of Fig.~\ref{fig:Seebeck}c),
which amounts to $\Delta \alpha / \alpha \simeq 5 \%$,
is within the uncertainty of this composite measurement; see Methods.)
%
%which involves taking the ratio of two separate quantities, $S$ and $\rho$,
%given that $S$ involves two types of contact -- electrical and thermal --  effective position along the length of the sample will in general be slightly different -- although measured from the same gold pads,
%resulting in an uncertainty on $S$ given roughly by contact width over contact separation,
%so about 10\%).
%}
%
This reveals that the CuO chains have no thermoelectric response.
This experimental observation suggests that the electronic dispersion of the CuO chains is perfectly particle-hole symmetric,
which remains to be understood.

This is very convenient for our study of nematicity, because in the Peltier coefficient
the background anisotropy from CuO chains is essentially zero.
So any anisotropy in $\alpha$ must be emerging spontaneously within the CuO$_2$ planes
(although we cannot rule out the possibility that the chain response changes at the onset of the charge density wave order below 140~K).
In Fig.~\ref{fig:Seebeck}c, we see that no such spontaneous anisotropy occurs at \Tstar.
So here again, now in the longitudinal Peltier coefficient, we detect no nematicity upon entry into the pseudogap phase.
We do observe a large anisotropy below 100~K or so, which we discuss in the next section.

%From $T=300$~K down to $T\approx 100$~K, the Peltier coefficient is basically isotropic $\Delta \alpha = \alpha_{\rm a} - \alpha_{\rm b} = S_{\rm a} / \rho_{\rm a} - S_{\rm b}/\rho_{\rm b} = 0$ (Fig.~\ref{fig:Seebeck}b,c), seamlessly going through \Tstar. First, this implies that the Peltier coefficient from the chains $\alpha_{\rm chains}$ is zero at all temperatures and that the anisotropy in Seebeck directly is given by the anisotropic ratio in the resistivity $S_{\rm b} = \frac{\rho_{\rm a}}{\rho_{\rm b}}S_{\rm a}$. Second, the Peltier coefficient demonstrate to only be sensitive to the physics of the CuO$_2$ planes, which makes it the best charge transport coefficient to look for nematicity. Having observed that $\Delta \alpha$ remains zero throughout \Tstar, this ultimately assesses that there is no onset of charge nematicity by entering the pseudogap phase.

% But let us try to dig a little further the Seebeck contribution from the Cu-O chains. In thermoelectricity, Seebeck coefficients do not add up, only Peltier coefficients do: $\alpha = \sigma S = S / \rho$. Therefore, one expects for $T>T*$:
% \begin{equation}
% \alpha_{\rm b} = \alpha_{\rm a} + \alpha_{\rm chains}
% \label{eq_peltier}
% \end{equation}

% Equation \ref{eq_peltier} can be rewritten:
% \begin{equation}
% S_{\rm b} = \frac{\sigma_{\rm a}S_{\rm a}+\sigma_{\rm chains}S_{\rm chains}}{\sigma_{\rm a}+\sigma_{\rm chains}}
% \label{eq_Sb}
%  \end{equation}

% >>>>>>>>>>>>>>>>>>>>>>>>>>>>>>>>>>>>>>>>>>>>>>>>>>>>>>>>>>>>>>>>>>>>>>>>>>>>>>>>>>>>>>>>>>>>>>>>>
\section{DISCUSSION}

%\textbf{CuO chains}. Do we want to discuss $T^2$ resistivity and $\alpha_{\rm chains} = 0$?

\textbf{Anisotropy in the Nernst coefficient}.
Our finding of no nematicity at \Tstar~deduced separately from two longitudinal transport coefficients, $\rho$ and $\alpha$, seems to contradict the prior finding of nematicity at \Tstar~reported by Daou \textit{et al.}~\cite{Daou2010Broken} deduced from the Nernst coefficient (a subset of the present authors published this earlier work).
There is in fact no contradiction between the absence of anisotropy in the longitudinal coefficients and the 
presence of a small anisotropy in the Nernst coefficient. 
The apparent nematicity at \Tstar~in the Nernst anisotropy is a small artifact of the composite nature of that coefficient.
Indeed, as seen from Eq.~1, the Nernst anisotropy difference
$(\nu_{\rm a} - \nu_{\rm b}) / T$
can in principle come from two terms, the first involving
$(\alpha_{\rm a} - \alpha_{\rm b}) / T$
and the second involving
$(\rho_{\rm a} - \rho_{\rm b})$.
While we now know that the first term is essentially zero between 300~K and 140~K (Fig.~\ref{fig:Seebeck}c),
the second term is nonzero, and it will change across \Tstar~$= 220$~K,
as can be seen by inspection of the resistivity data in Fig.~\ref{fig:resistivity}a.
The Nernst coefficient will therefore pick up that slight change across \Tstar.
%which does not reflect any true nematicity, as we now know.

\begin{figure}[t!]
\centering
\includegraphics[width=0.46\textwidth]{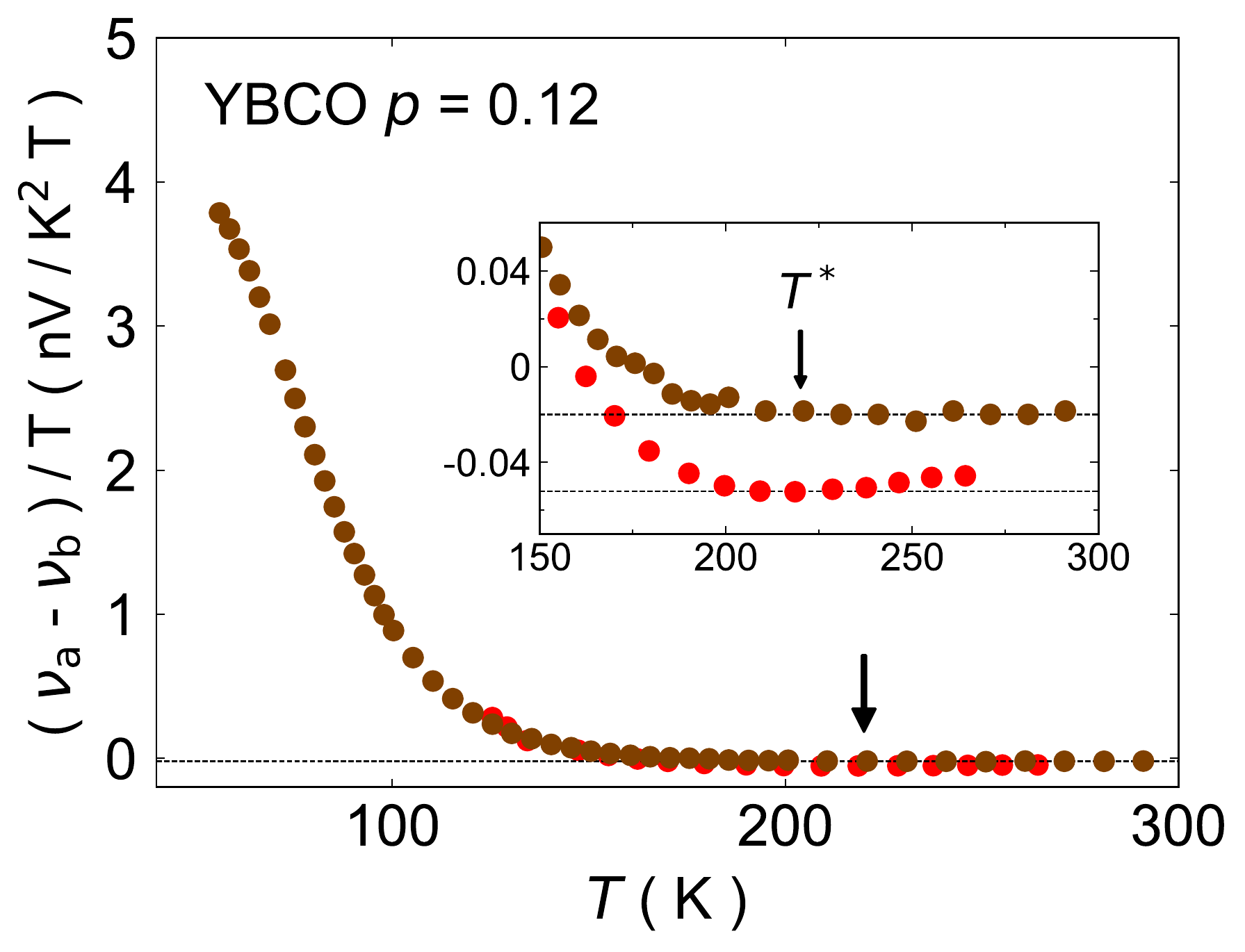}
\caption{
Anisotropy in the Nernst coefficient $\nu$, plotted as $(\nu_{\rm a} - \nu_{\rm b}) / T$ vs $T$ for our YBCO $p=0.12$ sample, at $H=16$~T (full brown circles).
Corresponding data reported by Daou \textit{et al.}~\cite{Daou2010Broken} are reproduced here in red.
The inset shows that the rise in the Nernst anisotropy starts roughly at the same temperature \Tstar 
in the two studies.
}
\label{fig:nernst}
\end{figure}

%Our transport results seem to contradict the onset of the anisotropy in the Nernst effect at \Tstar reported by Daou \textit{et al.}~\cite{Daou2010Broken}.
%
During the same two-stage experiment used to measure the anisotropy in the resistivity and the Seebeck coefficient,
we also measured the Nernst coefficient.
%
%On our current sample of YBCO with $p = 0.12$, we also measured the Nernst coefficient.
The resulting data are plotted in Fig.~\ref{fig:nernst} as the difference
%$\Delta\nu /T = (\nu_{\rm a} - \nu_{\rm b}) / T$ vs $T$.
%
$(\nu_{\rm a} - \nu_{\rm b}) / T$ vs $T$.
These are consistent with the corresponding data reported by Daou \textit{et al.}~\cite{Daou2010Broken}.
By zooming on the region at high temperature (inset), we see that the Nernst anisotropy starts to grow below $T \simeq 200$~K~$\simeq$~\Tstar.
This growth is very slow, as by 140~K the Nernst anisotropy has only reached 1~\% of its full value at low temperature.
It is this very slight additional anisotropy in $(\nu_{\rm a} - \nu_{\rm b}) / T$ vs $T$,
%completely absent from the longitudinal Peltier coefficient,
whose anisotropy $(\alpha_{\rm a} - \alpha_{\rm b}) / T$ is essentially zero from 300~K down to 140~K (Fig.~\ref{fig:Seebeck}c),
which led Daou \textit{et al.} to conclude that nematicity starts at \Tstar.
As we now know, it does not reflect any true nematicity.

%#################### Figure 4 ####################%
%##################################################%

\textbf{Nematicity associated with CDW ordering}.
Of course, the 100-fold growth in the Nernst anisotropy that takes place between 150~K and 50~K
does reflect a true additional anisotropy, in a composite way.
For a direct measure of nematicity, it is best to look at the Peltier anisotropy, because it is a longitudinal coefficient and it is essentially zero at high temperature.
In Fig.~\ref{fig:Seebeck}c, we see that
$(\alpha_{\rm b} - \alpha_{\rm a}) / T$ starts to rise below 140~K,
and this growth is smooth and continuous down to at least 35~K.
As done previously for the Nernst anisotropy~\cite{cyr-choiniere_two_2015},
we attribute the emergence of the Peltier anisotropy to the development of charge-density-wave (CDW) order.
At $H = 16$~T, the charge ordering process is known from x-ray studies to occur in two stages~\cite{hucker2014,blanco-canosa2014,gerber_2015,chang_magnetic_2016}:
first, short-range two-dimensional CDW correlations start to grow below $T_{\rm CDW1} = 140\pm10$~K~\cite{chang_direct_2012},
then long-range three-dimensional charge order sets in at $T_{\rm CDW2} = 47\pm1$~K~\cite{chang_magnetic_2016}.
The latter order is unidirectional, so it breaks rotational symmetry, as well as translational symmetry.
It is clear from Fig.~\ref{fig:Seebeck}c that the growth in Peltier anisotropy occurs in tandem with the CDW ordering process.
If one considers that long-range translational symmetry is only broken at $T_{\rm CDW2}$,
then the region between $T_{\rm CDW2}$ and $T_{\rm CDW1}$ would be a nematic phase,
where rotational symmetry is spontaneously broken in the electronic fluid of the CuO$_2$ planes,
in agreement with NMR studies~\cite{Wu2015Incipient}.

In a nutshell, our controlled measurements of longitudinal transport conducted on a single YBCO sample for both current directions ($a$ and $b$) allow us to show that previous signatures of nematicity based on Nernst measurements \cite{Daou2010Broken,cyr-choiniere_two_2015} do not onset at \Tstar, but at significantly lower temperature. 
Although we can only claim that electron fluid nematicity below \Tstar is unobservable within our resolution, 
the sharp signature reported at \Tstar in torque experiments~\cite{Sato2017Thermodynamic} remains, {\it a priori}, 
incompatible with the absence of charge nematicity in longitudinal transport coefficients. 
Further investigations will be needed to solve this apparent contradiction.

% >>>>>>>>>>>>>>>>>>>>>>>>>>>>>>>>>>>>>>>>>>>>>>>>>>>>>>>>>>>>>>>>>>>>>>>>>>>>>>>>>>>>>>>>>>>>>>>>>
\section{SUMMARY}

In summary, we have measured the in-plane anisotropy of the longitudinal charge transport coefficients of YBCO at a doping $p=0.12$
--- the resistivity and the Seebeck coefficient.
Upon cooling through the pseudogap temperature \Tstar,
we observe no additional in-plane anisotropy in the resistivity $\rho(T)$,
beyond the background anisotropy due to the CuO chains, whose resistivity obeys a $T^2$ dependence that can therefore be readily subtracted.
This shows that the pseudogap phase of cuprates is not nematic.
The Peltier coefficient $\alpha = S / \rho$ appears to have, surprisingly, no contribution from the CuO chains along the $b$ axis. This makes $\alpha$ a most ideal probe for studying nematicity in the CuO$_2$ planes of YBCO.
We find that the Peltier coefficient remains isotropic across \Tstar,
which confirms the absence of charge nematicity at the onset of the pseudogap phase.

However, a large anisotropy in the Peltier coefficient does emerge in tandem with the growth in the short-range charge density wave
correlations starting below $T_{\rm CDW1} = 140$~K.
This anisotropy continues to grow down to at least $T= 35$~K, below the onset of long-range charge density wave order at $T_{\rm CDW2}=47$~K (in $H=16$~T).
Since translational symmetry is only truly broken, on a long length scale, at $T_{\rm CDW2}$,
we infer that the regime between $T_{\rm CDW2}$ and $T_{\rm CDW1}$ may be a nematic phase.
In the absence of nematicity at \Tstar, the anisotropy previously detected by torque magnetometry at \Tstar\cite{Sato2017Thermodynamic} must find a different origin.

% % %-----------------------------------------%
% % %############# Acknowledgments ###########%
% % %-----------------------------------------%

\section*{Acknowledgements}

We thank S.~Fortier for his assistance with the experiments. L.T. acknowledges support from the Canadian Institute for Advanced Research (CIFAR) as a CIFAR Fellow and funding from the Institut Quantique, the Natural Sciences and Engineering Research Council of Canada (PIN:123817),
the Fonds de Recherche du Qu\'ebec -- Nature et Technologies (FRQNT), the Canada Foundation for Innovation (CFI), and a Canada Research Chair.

% %-----------------------------------------%
% %############### Biblio ##################%
% %-----------------------------------------%


\end{document}